\documentclass[aps,prl,twocolumn,superscriptaddress]{revtex4}
\usepackage{graphicx}
\usepackage{verbatim}

\newcommand{\req}[1]{(\ref{#1})}
\newcommand{\be}{\begin{equation}}
\newcommand{\ee}{\end{equation}}
\newcommand{\bea}{\begin{eqnarray}}
\newcommand{\eea}{\end{eqnarray}}

\newcommand{\id}{1\!\!1}
\newcommand{\bra}{\langle}
\newcommand{\ket}{\rangle}

\newcommand{\avg}[1]{\langle{#1}\rangle}

\newcommand{\BE}{\begin{eqnarray}}
\newcommand{\EE}{\end{eqnarray}}
\newcommand{\BEn}{\begin{eqnarray*}}
\newcommand{\EEn}{\end{eqnarray*}}
\newcommand{\barr}{\begin{array}}
\newcommand{\earr}{\end{array}}

\newcommand{\bit}{\begin{itemize}}      
\newcommand{\eit}{\end{itemize}}
\newcommand{\bc}{\begin{center}}
\newcommand{\ec}{\end{center}}
\newcommand{\ben}{\begin{enumerate}}    
\newcommand{\een}{\end{enumerate}}

\newcommand{\sign}{\mbox{sgn}}

\begin{document}

\title{Price return auto-correlation and predictability in agent-based models of financial markets}
\author{Damien Challet}
\affiliation{Nomura Centre for Quantitative Finance, Mathematical
  Institute, Oxford University, 24--29 St Giles', Oxford, United Kingdom}
\affiliation{The Rudolf Peierls Centre for Theoretical Physics, Oxford University, 1--3 Keble Road,
  Oxford, OX1 3NP, United Kingdom}
\author{Tobias Galla}
\affiliation{The Rudolf Peierls Centre for Theoretical Physics, Oxford University, 1--3 Keble Road,
  Oxford, OX1 3NP, United Kingdom}
\affiliation{International Center for Theoretical Physics, Strada Costiera 11
, 34014 Trieste, Italy}
\affiliation{Istituto Nazionale per la Fisica della Materia (INFM), Trieste-SIS
SA Unit, V. Beirut 2-4, 34014 Trieste, Italy}

%\address{}
\date{\today}

%%%%%%%%%%%%%%%%%%%%%%%%%%%%%%%%%%%%%%%%%%%%%%%%%%%%%%%%%%%%%%%%%%%%%%%%%%%%%

\begin{abstract}
We demonstrate that minority mechanisms
arise in the dynamics of markets because of price impact; accordingly the relative importance of minority and delayed majority mechanisms depends on the frequency of trading. We then use mixed majority/minority games to illustrate that a
vanishing price return auto-correlation function does not necessarily
imply market efficiency. On the contrary, we stress the difference
between correlations measured conditionally and unconditionally on
external patterns.

\end{abstract}
\maketitle

Whether financial markets are predictable or not is a debate whose
origin can be traced back to Bachelier's hypothesis that prices follow
a random walk \cite{Bachelier}. Later work, in particular by Fama
\cite{Fama} and Samuelson \cite{Samuelson} aimed at proving
mathematically that markets are efficient, i.e. unpredictable,
grounding their theories on perfect rationality. This is indeed the
simplest view of a financial market, and a very convenient one. There
are however good reasons to believe that markets are not perfectly
efficient. The well-known January effect is a systematic deviation
from efficiency and has been observed in real markets for many
years, but is reportedly disappearing; other examples include the predictive power of moving averages
\cite{QFFX}. According to more recent theoretical studies, the
cost of acquiring information \cite{ZMEM} or the presence of noise
traders \cite{Shleifer} prohibits perfect efficiency, as perfect
information can only be achieved at an infinite cost, or by accepting
a considerable risk; this leads to the hypothesis of marginally
efficient markets, where most of the easily detectable predictability
is removed by traders.

Market efficiency is often illustrated by the fact that the price
return auto-correlation function is essentially zero. Denoting the
log-price at time $t$ by $p(t)$, the price return is defined as $r(t)=
p(t+1)-p(t)$, and assuming that $r(t)$ is a stationary process, its
auto-correlation reads
\be\label{CTtau}
C(\tau)=\left(\avg{r(t+\tau)r(t)}-\avg{r(t)}^2\right)/\avg{r(t)^2}.
\ee

Here $\avg{\cdots}$ stands for averages over time; note that we
normalise by the average magnitude of returns. The units of time are
arbitrary in this paper. If $C(\tau)\ne 0$, statistically significant
predictions of future price changes are possible on time-scales
$\tau$, that is, the knowledge of $r(t)$ allows to make probabilistic
statements about $r(t+\tau)$. The analysis of real market data shows
that $C(\tau)\approx 0$ typically for $\tau$ larger than a few minutes
\cite{LoBook,Daco,BouchaudPottersBook}. However, the existing
correlations on shorter time-scales are not exploitable in reality because of transaction costs \cite{BouchaudPottersBook}.

This apparent efficiency however does not deter some practitioners
from making statistically abnormal profits. In this paper we argue
that the correlation function as defined in Eq. \req{CTtau} is a poor 
measure of predictability. It is indeed
an {\em unconditional} measure of correlation and does not
differentiate between technical analysis patterns, states of the market or of the economy. Finding and exploiting relevant information patterns generates arbitrage opportunities. As a consequence, correlation functions conditional on these patterns have to be used.

In the following we will use agent-based
market models in which patterns play an essential role, to illustrate the differences
between measures of correlation in price time series which are
conditional or unconditional on these patterns.

\section{Patterns}

We now formalise the notions of conditional and unconditional
averages as introduced above. Assume for example that there are $P$
patterns $\mu=1,\cdots,P$ which are relevant (or believed to be
relevant) for the behaviour of the market. The pattern or state of the
market at time $t$ will be denoted by $\mu(t)$.
Furthermore we will assume for simplicity that all
patterns occur with equal probability $1/P$. Such patterns can then
potentially be used in order to predict future price changes and in
this sense money making consists essentially in identifying a suitable
set of patterns and in exploiting the resulting predictability
conditional on a given pattern, if it is present.

The average return conditional on $\mu$ will be denoted by
$\avg{r|\mu}$ in the following, that is the time average
$<\cdots|\mu>$ is only performed over instances $t$ for which
$\mu(t)=\mu$. If at time $t$, $\mu(t)=\mu$, $r(t)$ is statistically predictable if $|\avg{r|\mu}|>0$; in addition, if  $|\avg{r|\mu}|>c$ where $c$ denotes the 
transaction cost, the price return predictability given $\mu$ can
be exploited
\footnote{In practice of course one should also take into account the
risk, i.e. fluctuations of $r$ conditional on $\mu$.}. We will neglect
transaction costs in our discussion. Given a set of $P$ states,
the predictability, $H$, can now be defined as follows
\footnote{In principle, a better measure of the predictability would
be $\frac{1}{P}\sum_{\mu=1}^P |\avg{r|\mu}|$. However the definition
used here is more convenient.} 
\be\label{H}
H=\frac{1}{P}\sum_{\mu=1}^P \avg{r|\mu}^2.  
\ee 
Note that $H$ measures the average predictability of the next price return. This means that the time horizon over which this predictability can be exploited is fixed, and equal to the time interval over which returns are defined. In addition, $H$ fails to measure predictability associated with oscillatory behaviour. The latter is captured instead by the conditional price return auto-correlation function $K(\tau)$ defined as follows

\be\label{eq:Kdef} 
K(\tau)=\left(\frac{1}{P}\sum_{\mu=1}^P \avg{r(t)r(t+\tau)|\mu}-H\right)/\avg{r(t)^2}. 
\ee
Here $\avg{r(t)r(t+\tau)|\mu}$ measures the correlation of price
returns between two occurrences of a given pattern $\mu$, i.e. we
impose $\mu(t)=\mu(t+\tau)=\mu$ when taking the above average.
Although $H$ is a measure for conditional predictability, $K$ can
provide additional predictability, as we will see agent based market
models can display regimes in which $H$ vanishes, but where at the
same time $K$ is non-zero. Secondly, $K$ is of interest when the
signal-to-noise (or Sharpe) ratio $H/(\avg{r^2}-H)$ is small, that is, when it is
likely that for a fixed $\mu$, $A$ changes sign between two
occurences of that pattern.

The aim of the present paper is not to identify sets of predictability
patterns in real market time series (see e.g. \cite{MarsiliClust} for a method of extracting patterns from real market data). We shall rather illustrate
the difference between conditional and unconditional measures of
predictability in agent-based models and will accordingly focus on
models which are able to capture the relationship between
predictability and price return auto-correlation. Agent-based models
of financial market are appealing because they make it possible to
study the relationship between the behaviour of the agents and the
statistical properties of the resulting price time series. Early studies \cite{SantaFe,Takayasu,Lux,ContBouch}
showed that such models are able to reproduce some typical market-like
data, so-called stylized facts.  Recently, modified Minority Games
\cite{CZ97} where the agents are allowed not to play if they do not
believe to be able to make money, exhibit similar features \cite{SZ99,J99,J00,CMZ01,CM03}. Such models are on the
borderline of being exactly solvable by methods of statistical
physics, while showing non-trivial realistic behaviour, and are hence
of interest to both economists and statistical physicists. Volatility
clustering obtained in agent based models has been related to strategy
switching \cite{BouchMG}, and large price returns to market efficiency
\cite{CM03}. 

\section{Price evolution, minority and majority games} 

Before turning to specific agent-based models we first define a
price evolution mechanism and  reward structure. Let us concentrate on one given trader who opens a position $a$ at time $t$, which can be either long ($a=1$), or short ($a=-1$). After some waiting time, say at time $t'>t$, the same trader closes his position. His gain is therefore given by
\be
a[p(t'+1)-p(t+1)],
\ee
as $p(t+1)$ and $p(t'+1)$ are the prices at which the transactions take place.
We assume that the excess demand at times $t$ and $t'$ are given by $A(t)$ (which includes $a$), and $A(t')$ (which includes $-a$), respectively. It is reasonable to suppose that the
excess demand $A(t)$ has a price impact of $f[A(t)]$, i.e. that
$p(t+1)=p(t)+f[A(t)]$. For convenience, we will consider a simple
linear price impact function $f(x)=x/\lambda$, where $\lambda$ is the
liquidity \cite{Farmer}.  We will set $\lambda=1$ without loss of
generality. Therefore, the gain of this trader in this
round trip reads 
\bea\label{realpayoff}
a[p(t'+1)-p(t+1)]&=&-(-a)A(t')-aA(t)\nonumber\\
&&+a[p(t')-p(t)].
\eea 
The first two terms on the right hand side are typical of a
minority game payoff, in which this trader is rewarded if his action
$a$ is that of the minority at that time, i.e., the payoff $-aA(t)$
is positive whenever $a$ and $A(t)$ have opposite signs. The last
term represents the payoff that could be obtained if there were no
price impact; its nature is that of a cumulative and delayed majority
game. In other words, because of price impact, the market is a minority game
when positions are opened or closed. When a position is held, the market is a retarded majority game, as it is
favourable to hold a long position whenever the price is increasing
($A>0$) and vice versa. 
%This is roughly equivalent to the
%\$-game of Ref. \cite{BouchaudGiardina,Sornette}; 
%since the market
%impact is absent from this payoff we shall keep calling this mechanism
%a retarded majority game. 
As a consequence, the relative importance
of minority and majority mechanisms depends on the frequency of
trading. The Minority Game \cite{CZ97} imposes to trade at each time
step, and hence focuses on price impact.

The first paper to apply the minority game framework to a two
time-step payoff is \cite{SZ99}, where the payoff at time $t$ is
$a[N_in(t)-N_in(t-1)]$, $N_in(t)$ being the number of people with a
long position at time $t$ (no short position is allowed) and is
similar to our price $p(t)$; therefore, this payoff can be translated
into $a(t)A(t+1)$.  The idea is to reward people for anticipating the
crowd. Later work \cite{BouchaudGiardina,Sornette} explicitly rewarded
the agents with $a(t)A(t+1)$, allowing also for short selling; games
with this type of payoff are often referred to as '\$-games'.
%this payoff
%implicitly assumes that the position is opened at time $t$ and closed at
%time $t+1$, and that there is no price impact in the sense of Eq
%\req{realpayoff}. 
Eq. \req{realpayoff} therefore extends previous work in two
crucial aspects: first by adding the holding of a position over several time steps and
secondly by considering price impact. The former is reflected by
payoffs $a(t)A(t)$, $a(t)A(t+1),\cdots,a(t)A(t'-1)$, i.e. a total
gain of $a(t)\left[A(t'-1)+\dots+A(t)\right]=a(t)\left[p(t')-p(t)\right]$ while
holding the position, and price impact leads to the terms $-a(t)A(t)$
and $-a(t')A(t')$.  It should be noted that equation \req{realpayoff}
states that the real gain from a round-trip is composed of a sum of sequential payoffs; therefore it also contains a description
of how profits and losses (P\&L) and trading strategy gains are to be updated in real time.

Exact solutions for models with two-time step payoffs are currently
unavailable, but sometimes an understanding of their qualitative
properties is possible (e.g. \cite{BouchaudGiardina}).  Although
realistic speculative trading inevitably involves two or more
time-steps, the perception of price dynamics by the agents can be
captured by models involving only one-step payoffs: Ref. \cite{M01}
derives the \$-game (which does not include price impact) rigorously
and argues that different players can have different types of one-step
payoffs depending on their {\em a priori} beliefs on the market: it
concludes that fundamentalists, or value traders, believe that the
market is a minority game, while trend-followers believe it to be a
majority game. Thus, it is possible to consider a simple market with
constant proportions of fundamentalists and trend-followers,
respectively, that is, a mixed population of minority and majority
players. An analysis along the lines of
\cite{M01} can be applied to the last term of Eq \req{realpayoff} of course, taking
into account possible different beliefs of the agents on the future
behaviour of the market. This procedure would not, however, lead to
one-step payoffs as the two market impact terms would not disappear.

Since mixed minority/majority game models are exactly dynamically solvable
\cite{M01,Andrea,MarsiliMaj}, they will be the focus of this paper (see \cite{ChalletPatterns} for a model of speculation based on Eq \req{realpayoff}).

\section{Minority/majority games: definition}
\label{defMG}

A minority/majority game describes an ensemble of $N$ agents, each of
whom at every time-step has to take one of two possible actions,
labelled by $-1$ and $+1$. We will denote the choice of agent $i$ at
time $t$ by $a_i(t)$. The excess demand is $A(t)=\sum_{j=1}^N
a_j(t)$. The payoff given to agent $i$ is \be\label{minmaj} \phi_i
a_i(t)A(t), \ee where $\phi_i=1$ if agent $i$ is a trend follower,
thus prefers to be in the majority; conversely, if $\phi_i=-1$, agent
$i$ is a fundamentalist, and plays a minority game: he is rewarded if
his action $a_i(t)$ has the opposite sign of the total bid $A(t)$
\cite{M01}. We assume that there is a constant fraction $\theta$ of
minority players, so that choosing $\theta\in[0,1]$ allows to
interpolate between pure majority populations ($\theta=0$) and pure
minority games ($\theta=1$). We adopt the assumption of a linear price
impact function as above so that price returns are taken to be
proportional to the excess demand $A$.

\subsection{Games without patterns}\label{corrP=0}

As a warm-up we first compute the price return autocorrelation function of models without patterns as they are simple and analytically tractable. The adaptation abilities of the agents follow a simple
rule: agent $i$ computes the cumulative payoff,
i.e the score of action $+1$, denoted by $\Delta_i(t)$; at time $t$ his trading decision is probabilistic:
\be
P(a_i(t)=1)=\frac{1+\phi_i\tanh(\Gamma \Delta_i(t))}{2}. \ee
The scores themselves evolve according to \be\label{Deltai}
\Delta_i(t+1)=\Delta_i(t)+ A(t)/N.  
\ee
This defines a reinforcement learning known as the Logit model in economics
\cite{Logit} and corresponds to Boltzmann weights in statistical
physics \footnote{The original minority game was defined with
$\Gamma=\infty$. $\Gamma<\infty$ was first seen in \cite{Oxf1}.}. $\Gamma$ is a learning rate. Note
that if $\phi_i=1$ a large score $\Delta_i$ will lead to a preference
for action $a_i(t)=1$, while minority players ($\phi_i=-1$) will
predominantly play $a_i(t)=-1$ if they find a positive
$\Delta_i$. 

Minority games without patterns were introduced in \cite{MC00}, while
pattern-less mixed minority/majority games are first found in
\cite{M01}. If all the agents have the same initial score valuation
$\Delta_i(0)=\Delta(0)$, all the $\Delta_i$ are equal at all times,
hence one can drop the index $i$ in Eq. \req{Deltai}. $A(t)$ is then
the sum of $N$ independent identically distributed variables. In particular
$A(t)/N$ converges to $(1-2\theta)\tanh(\Gamma\Delta)$ for large
$N$. This means that in this limit the dynamics of $\Delta$ is well
described by \be\label{Eq:Delta}
\Delta(t+1)=\Delta(t)+(1-2\theta)\tanh(\Gamma \Delta(t)). \ee
\begin{figure}
\centerline{\includegraphics[width=0.4\textwidth]{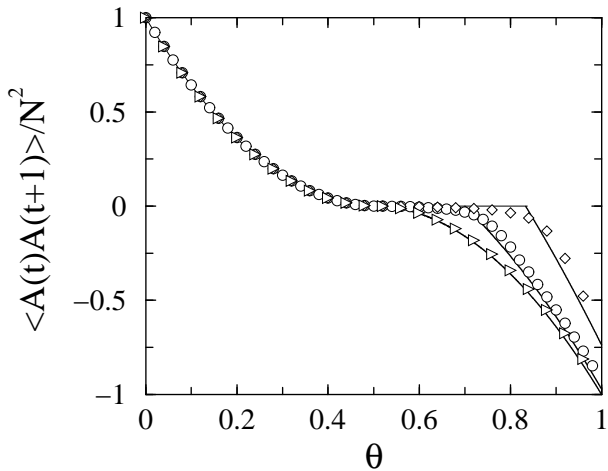}}
\caption{Autocorrelation function $\avg{A(t)A(t+1)}/N^2$ as a function of the fraction of minority players $\theta$ for $\Gamma=3$ (diamonds), $5$ (circles), and $100$ (triangles); $N=101$, 10000 iterations. Continuous lines are theoretical predictions.}
\label{C1P1}
\end{figure}

In the majority regime ($\theta<1/2$), one has $1-2\theta>0$ so that
$\Delta$ converges to $\pm\infty$: all the majority players agree on a
common decision and are happy, while the minority players agree on
the opposite decision and are equally happy: this is a Nash
equilibrium. $A$ is constant; more precisely,
$\frac{|A|}{N}=(1-2\theta)$, therefore
$\frac{\avg{A(t)A(t+1)}}{N^2}=(1-2\theta)^2$, which is confirmed by
simulations, see Fig. \ref{C1P1}.

The minority game regime ($\theta>1/2$) is slightly more complicated,
  as Eq. \req{Eq:Delta} has a unique fixed point which depends on
  $\Gamma(2\theta-1)$. If $\Gamma(2\theta-1)<2$, the fixed point
  $\Delta^*=0$ is stable and unique, leading to $\avg{A}=0$
  \cite{M01}; this also implies that $\avg{A(t)A(t+1)}=0$. On the
  other hand, if $\Gamma(2\theta-1)>2$, $\Delta=0$ becomes unstable,
  and a period two dynamics emerges \cite{M01}. The oscillation
  amplitude of $\Delta(t)=(-1)^t\Delta^*$ is determined by \be
  \Delta^*=\frac{(2\theta-1)}{2}\,\tanh(\Gamma \Delta^*), \ee which has
  of course two solutions of opposite signs. Neglecting fluctuations
  around $\Delta^*$, the resulting one-step correlation function is
  \be\label{EqAtAt+1Min}
  \frac{\avg{A(t)A(t+1)}}{N^2}\simeq-[(2\theta-1)\tanh(\Gamma
  \Delta^*)]^2.  \ee In Fig. \ref{C1P1} we show $\avg{A(t)A(t+1)}/N^2$
  computed from numerical simulations of the mixed minority/majority
  game without patterns, and compare the results with the simple
  approximation of Eq. \req{EqAtAt+1Min}. We find that both are in good
  agreement. As discussed in Refs. \cite{Farmer,M01}, the sign of
  $\avg{A(t)A(t+1)}$ reveals the minority/majority nature of the
  market, that is, the fraction of value investors and trend followers: if the correlation function is positive majority players
  dominate and vice versa. As the outcome of the game in the majority regime is constant, its auto-correlation $C(1)=0$. The case of the minority regime is more interesting: as $\avg{A}=0$, the agents cannot predict the next outcome; in addition, if $\Gamma(2\theta-1)\le2$, its correlation function $C(1)=0$, hence, the game is really unpredictable; if $\Gamma(2\theta-1)>2$ and its correlation function $C(1)<0$ measures predictability that cannot be exploited by the agents defined above, but could be by more sophisticated agents that would have strategies taking into account correlations.

\subsection{Games with patterns}

When the agents believe that the next outcome of the game depends on
some information such as economic forecast, the price history or the
number of sun spots, their trading actions
at time $t$ will depend on the pattern $\mu(t)$. Accordingly the
future evolution of the price itself will be correlated with these
pieces of information, hence the need for conditional averages.

In the original Minority Game \cite{CZ97}, the agents are given pieces of information about the last price returns. The agents in this game are therefore technical analysists. More specifically, the public pattern $\mu$ at time $t$
consists in the signs of the past $M$ price returns; therefore in the original Minority Game, $\mu(t)$ can be seen as a
binary string of length $M$ ($\mu(t)=1,\cdots,2^M$).
 Interestingly, Ref.
\cite{Cavagna} showed that the averaged price fluctuations are mostly
unchanged if $\mu$ is chosen randomly from $\{1,\cdots,2^M\}$. On the other hand crashes extending over several time-steps only occur if the patterns encode the real price history  \cite{Jcrash}.

Here we will only consider random patterns. Since the agents act conditionally
upon a given pattern $\mu(t)$ at time $t$ and are forced
to trade at each time step, a strategy $a$ is a
function which maps the pattern $\mu$ onto a trading action
$a^\mu=+1$ or $-1$ for each pattern $\mu=1,\cdots,P$. Strategies can hence be viewed
as $P$-dimensional vectors with binary entries. We will assume that
each agent holds two such strategies, which are assigned at random at
the start of the game and are then kept fixed \footnote{Minority games with larger pools of strategies per agent show qualitatively similar
behaviour.}. 
%Then, at each time-step
%every agent uses the one which he thinks has performed best so
%far. 
Following the now standard formalism, we denote the the
strategy used by agent $i$ at time step $t$ by $s_i(t)=\pm 1$, so that
upon presentation of pattern $\mu(t)$ at $t$ his action will read
$a_i(t)=a_{s_i(t)}^{\mu(t)}$. This can be written as
$a_{i}(t)=\omega_i^{\mu(t)}+s_i(t)\xi_i^{\mu(t)}$ where
$\omega_i^\mu=\frac{1}{2}(a_{i,1}^\mu+a_{i,-1}^\mu)$ and
$\xi_i^\mu=\frac{1}{2}(a_{i,1}^\mu-a_{i,-1}^\mu)$. The excess demand at
time $t$ can then be decomposed as follows:
\be\label{decompositon}
A(t)=\Omega^{\mu(t)}+\sum_{i=1}^N\xi_{i}^{\mu(t)}s_i(t), \ee where
$\Omega^\mu=\sum_{i=1}^N\omega_i^\mu$.

In order to decide which of his two strategies to play each agents
assigns a virtual score to each of his strategies, based on the payoff
he would have obtained had he always played this particular
strategy. In the case where the agents have two strategies, only the
score difference $q_i$ matters, and evolves according to 
\be\label{qit}
q_i(t+1)=q_i(t)+2\phi_i\xi_i^{\mu(t)}A(t)/P. 
\ee 
As in Eq. \req{minmaj}, $\phi_i=\pm 1$ describes the nature of the trader. At time $t$, agent $i$  then plays strategy $s_i(t)=\sign\ q_i(t)$.

This is the so-called online version of the
game, where the agents adapt their scores after every round of the
game. It was observed that the qualitative behaviour of the model does not
change much if one allows the agents to switch strategies only after a
large number of rounds have been played \cite{Moro1}. This effectively amounts to
sampling all the patterns  $\mu=1,\cdots,P$ and
leads to the so-called batch version of the game: \be\label{eq:batchupdate}
q_i(t+1)=q_i(t)+\frac{2}{P}\phi_i\sum_{\mu=1}^{P} \left[\xi_i^\mu\left
( \Omega^{\mu}+\sum_{j=1}^N\xi_{j}^{\mu}s_j(t)\right)\right], \ee
which is much less demanding to study analytically than the online
game \cite{CoolenBatch,CoolenOnline}. Note that the time $t$ in Eq. \req{eq:batchupdate} is not the same as that of Eq. \req{qit}, as it has been rescaled by a factor $P$ to take into account the fact that the agents are allowed to switch strategies once every $P$ 'online' time-steps.

The statistical mechanics analysis of the minority game has revealed a
phase transition between a predictable and an unpredictable phase. The
relevant control parameter is the ratio $\alpha=P/N$ between the
number of different patterns $P$ and the number of players $N$. For
$\alpha<\alpha_c\approx 0.33$ the market is unpredictable so that $H$
as defined above vanishes. This is illustrated in the upper panels of
Figs \ref{C1} and \ref{xi_1}. For $\alpha>\alpha_c$ one finds $H>0$ so that the knownledge of $\mu(t)$ is enough to predict the next price return \cite{CMZ99}.

\subsubsection{Online games}

The setup of the online game makes it possible to compare the price return
auto-correlation function $C(\tau)$, with $K(\tau)$, the conditional price return auto-correlation function as defined in Eq. (\ref{eq:Kdef}). 

\begin{figure}
\centerline{\includegraphics[width=0.4\textwidth]{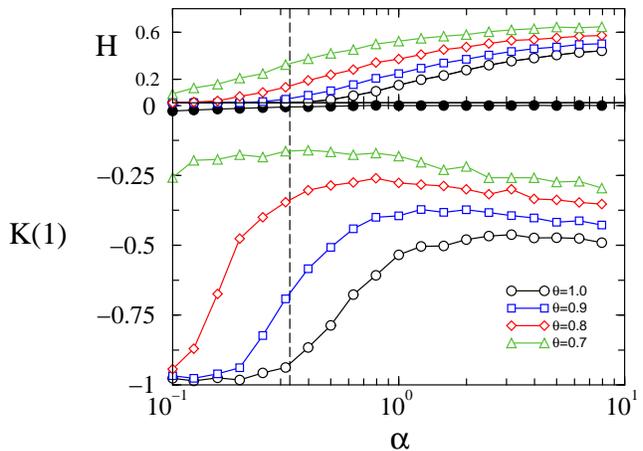}}
\caption{Predictability $H$ (upper panel), and conditional one-step autocorrelation function $K(1)$ vs $\alpha=P/N$ for the online game and different values of $\theta$ (open symbols). The unconditional one-step auto-correlation $C(1)$ is close to zero for all values of $\alpha$ and $\theta$, and is shown in the lower panel for $\theta=1$ (full circles). The vertical dashed
line marks the transition between unpredictable and predictable
states for the pure minority game ($\theta=1$). Simulations were performed on systems of $N=500$ agents, run for $10^6$ on-line time-steps, results are averaged over 20 random assignments of the strategies.}
\label{C1}
\end{figure}

\begin{figure}
\centerline{\includegraphics[width=0.4\textwidth]{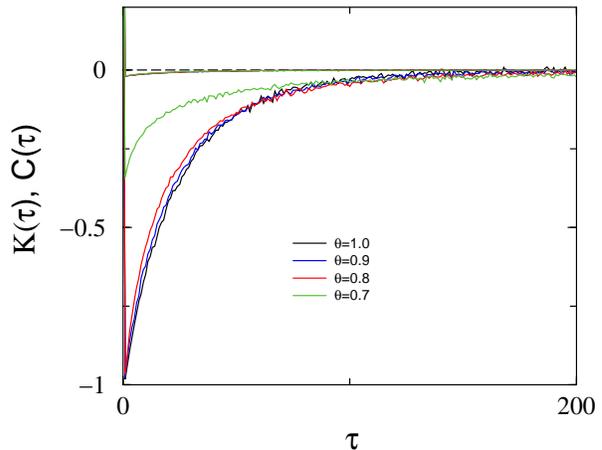}
}
\caption{
Unconditional and conditional price return autocorrelation functions
$C$ and $K$ vs time-lag $\tau$ for the online game at $\alpha=0.1$ and for different values of $\theta$. The upper curves are $C(\tau)$, the lower
$K(\tau)$. Results are from simulations of the online game with $N=500$ agents, run for $10^6$ time-steps and averaged over $10$ random assignments of the strategies.} \label{FigcorrP}
\end{figure}

\begin{figure}
\centerline{\includegraphics[width=0.4\textwidth]{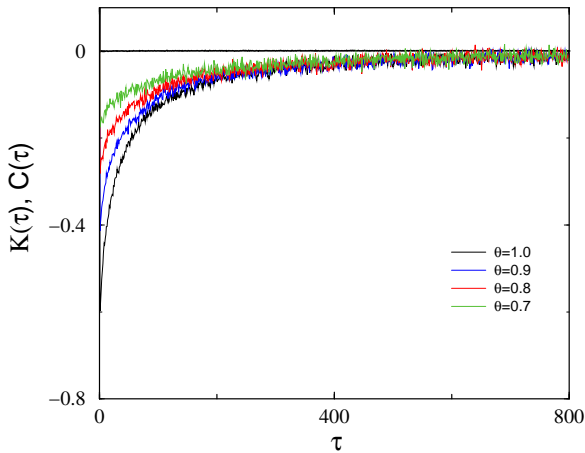}
}
\caption{
Conditional price return autocorrelation function $K$ vs time-lag
$\tau$ for the online game at $\alpha=0.75$ and for different values
of $\theta$. The unconditional correlation function $C(\tau)$ is
indistinguishable from zero on this scale. Results are from simulations of the online game with $N=300$ agents, run for $10^6$ time-steps and averaged over $10$ random assignments of the strategies } \label{FigcorrP2}
\end{figure}

Both auto-correlation functions are negative for $\tau=1$, as the minority game induces a mean-reverting process (\cite{M01,Mcorr}). The lower panel of Fig. \ref{C1} 
illustrates that the conditional correlation function $K(1)$ is much
larger than the unconditional one $C(1)$ for all values of $\alpha$. Note in particular that $C(1)$
is close to zero in the predictable phase ($H>0$) (and converges to 0 in the limit of infinite systems), whereas $K(1)\simeq -0.5$, confirming that unconditional measures fail to detect existing
arbitrage opportunities which can only be revealed by conditional
approaches. Fixing $P/N$ and plotting $C(\tau)$ and $K(\tau)$ as a
function of the time-lag $\tau$ confirms this statement  (see Figs.
\ref{FigcorrP} and \ref{FigcorrP2}); when $\theta$ decreases, the mean reversion of the price decreases because the fraction of minority players becomes less and less important; accordingly the amplitude of  $K$ decreases as well, and converges to 0 when $\theta=1$, since in that case the agents are all majority players, and only play one of their strategies, which does not produce any fluctuation of $A$ around $\avg{A|\mu}$.  When $H=0$, as $\alpha\to0$, $K(1)\to-1$ for sufficiently large $\theta$, meaning that for a given $\mu$, $A$ acquires an oscillatory behaviour. This type of predictability is not captured by $H$, but by $K(1)$, and, to a much lesser extent, by $C(1)$. 

\subsubsection{Batch games}

\begin{figure}
\centerline{\includegraphics[width=0.4\textwidth]{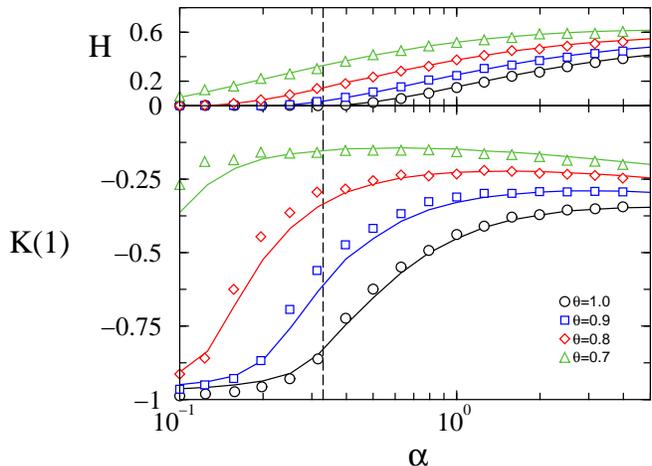}}
\caption{Predictability (upper panel) and one-step autocorrelation function $K(1)$ (lower panel) vs
$\alpha$ for the batch game at different values of $\theta$. Data
displayed as open symbols are obtained from numerical simulations of the batch process with
$N=300$ players, run for $100$ time-steps and averaged over $30$
realizations of the strategy assignments. The solid lines are results
from the Monte-Carlo integration of the self-consistent theory,
displayed for comparison ($60$ time-steps of the integration are performed, and results are averages over $2\cdot 10^5$ samples of the effective agent process). The vertical dashed line marks the transition between
unpredictable and predictable states for the case of the pure minority
game ($\theta=1$).}
\label{xi_1}
\end{figure}

\begin{figure}
\centerline{\includegraphics[width=0.4\textwidth]{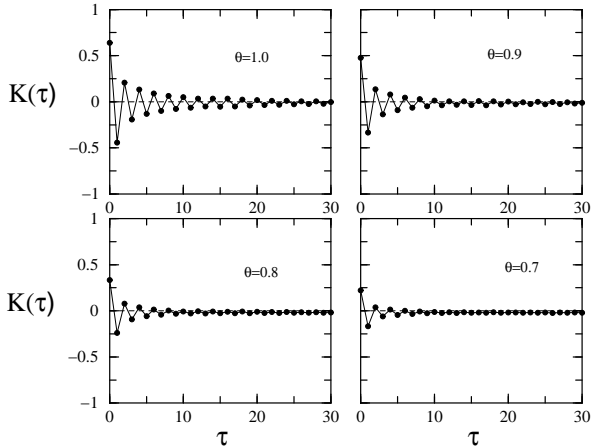}}
\caption{Autocorrelation function $K(\tau)$ vs $\tau$ for the batch 
Minority Game at $\alpha=P/N=1$ obtained by Monte-Carlo simulations of Eq. \req{eq:singleeffagent}, the iteration is performed for $60$ time-steps and sampled over $2\cdot 10^5$ single effective agent paths.}\label{corrfctbatch}
\end{figure}

The setup of the batch game allows only the measurement of $K(\tau)$, the conditional price return auto-correlation function (\ref{eq:Kdef}), because $A$ is replaced by its weighted sum over the patterns in Eq. \req{eq:batchupdate}. Fig. \ref{xi_1} shows that the behaviour of $K(1)$ as a function of $\alpha$ is qualitatively the same as in the online game (Fig. \ref{C1}). However, $K(\tau)$ itself is very different in batch games, as it oscillates as function of $\tau$ (Fig. \ref{corrfctbatch}). This is a signature of the mean-reverting process induced by the Minority Game players, and of the batch process, which updates the scores with respect to all the patterns at the same time.

The dynamics of the batch Minority Game was analyzed in Ref. \cite{CoolenBatch}, where it is proved that the $N$ coupled equations \req{eq:batchupdate} can be replaced, in the  limit of $N\to\infty$, $P\to\infty$ at fixed $\alpha=P/N$, by a single equation describing the
evolution of a representative agent. This result relies on the so-called generating functional analysis, a technique originally devised to study neural networks and other disordered systems \cite{Dominicis}. This method consists in averaging the dynamics over the disorder (i.e. over all possible strategy assignments) and is exact. The resulting equation reads 

\begin{equation}\label{eq:singleeffagent}
q(t+1)=q(t)+\alpha\phi\sum_{t^\prime\le t}(\id+G)^{-1}_{tt^\prime}\sign(q(t^\prime))+\theta(t)+\sqrt{\alpha}\eta(t),
\end{equation}
where $\phi=\pm 1$ again distinguishes between majority and minority
traders. While the original batch game is Markovian, this equation contains two
different types of memory: a retarded self-interaction term (the
second term in the above equation) relating to earlier times
$t^\prime\leq t$ and coloured Gaussian noise $\eta(t)$ with temporal
covariances
\begin{equation}
\bra \eta(t)\eta(t^\prime)\ket =
\bigg[(\id+G)^{-1}(E+C)(\id+G^T)^{-1}\bigg]_{tt^\prime}.
\end{equation}

The correlation function $C_{tt^\prime}=\bra
\sign(q(t))\sign(q(t^\prime))\ket$ and response function
$G_{tt^\prime}=\bra\frac{\partial
\sign(q(t))}{\partial\theta(t^\prime)}\ket$ to external perturbations
$\theta(t)$ are to be computed self-consistently as averages over the noise.
 $\id$ is the identity matrix, and
$E$ has all entries equal to one, $E_{tt^\prime}=1$.

Because of the non-Markovian nature of the single agent process and
the presence of coloured noise a full analytical solution of this
self-consistent problem is in general impossible. Assuming independence from initial conditions $q(0)$, 
the predictability $H$ can be computed exactly, but the calculation of
quantities like the volatility $\avg{A^2}$ or one-step correlation functions
requires the knowledge of both the persistent and transient parts of
the correlation and response functions. Neglecting correlations leads
to accurate approximations for the market volatility \cite{CoolenBatch}, but
unsurprisingly analogous approximations are inadequate for the
computation of price return correlation functions. One therefore has
to resort to a Monte-Carlo integration of Eq. (\ref{eq:singleeffagent}),
and as shown in Fig. \ref{xi_1} we find nearly perfect agreement with
simulations of the original batch process. The figure also illustrates
that the behaviour of the conditional one-step correlation function is
very similar in batch and online minority
games. Fig. \ref{corrfctbatch} reports the behaviour of the
conditional auto-correlation as a function of the time lag $\tau$ from Monte-Carlo simulations. One observes an oscillatory behaviour for small $\tau$, the oscillation
amplitude can be reasonably well fitted by
$\exp\left(-\tau^\beta/\kappa\right)$, with some constant
$\kappa$.

\section{Conclusions}

We have used agent-based models to illustrate the
 differences between conditional and unconditional
price return auto-correlation functions in financial markets. Unconditional indicators may be zero even though conditional measures reveal very large correlations. This emphasizes the inappropriateness of the use of purely unconditional auto-correlation functions for weak proofs of market efficiency. Therefore, a more correct statement about market efficiency would be that they are remarkably efficient {\em on average} (over patterns, over time), but not locally in time \cite{Wyart} or given relevant market states. 

As noted above, practitioners try to identify patterns $\mu$ in real markets and exploit resulting  predictability in order to make profit. The non-parametric
clustering method recently developed by Marsili for example
\cite{MarsiliClust} automatically extracts a set of patterns from market
data, which are at borderline of being predictive. It would be interesting
to compute price return auto-correlation functions, conditional on
patterns obtained by this method.
  
\section{Acknowledgements}
DC would like to thank Wadham College for support.  TG would like to
thank EPSRC for financial support under studentship 00309273 and
acknowledges the award of a Rhodes Scholarship and support by Balliol
College, Oxford and by the European Community's Human Potential
Programme under contract HPRN-CT-2002-00319, STIPCO. Discussions with
Jean-Philippe Bouchaud, Matteo Marsili and David Sherrington are
gratefully acknowledged.  \bibliographystyle{apsrev}
\bibliography{corr}

\begin{thebibliography}{40}
\expandafter\ifx\csname natexlab\endcsname\relax\def\natexlab#1{#1}\fi
\expandafter\ifx\csname bibnamefont\endcsname\relax
  \def\bibnamefont#1{#1}\fi
\expandafter\ifx\csname bibfnamefont\endcsname\relax
  \def\bibfnamefont#1{#1}\fi
\expandafter\ifx\csname citenamefont\endcsname\relax
  \def\citenamefont#1{#1}\fi
\expandafter\ifx\csname url\endcsname\relax
  \def\url#1{\texttt{#1}}\fi
\expandafter\ifx\csname urlprefix\endcsname\relax\def\urlprefix{URL }\fi
\providecommand{\bibinfo}[2]{#2}
\providecommand{\eprint}[2][]{\url{#2}}

\bibitem[{\citenamefont{Bachelier}(1900)}]{Bachelier}
\bibinfo{author}{\bibfnamefont{L.}~\bibnamefont{Bachelier}},
  \emph{\bibinfo{title}{Th\'eorie de la Sp\'eculation}} (\bibinfo{year}{1900}),
  vol.~\bibinfo{volume}{3} of \emph{\bibinfo{series}{Annales Scientifiques de
  l'Ecole Normale Superieure}}, pp. \bibinfo{pages}{21--86}.

\bibitem[{\citenamefont{Fama}(1970)}]{Fama}
\bibinfo{author}{\bibfnamefont{E.~F.} \bibnamefont{Fama}}, \bibinfo{journal}{J.
  of Fin.} \textbf{\bibinfo{volume}{25}}, \bibinfo{pages}{383}
  (\bibinfo{year}{1970}).

\bibitem[{\citenamefont{Samuelson}(1965)}]{Samuelson}
\bibinfo{author}{\bibfnamefont{P.~A.} \bibnamefont{Samuelson}},
  \bibinfo{journal}{Industrial Management Review} \textbf{\bibinfo{volume}{6}},
  \bibinfo{pages}{41} (\bibinfo{year}{1965}).

\bibitem[{\citenamefont{James}(2003)}]{QFFX}
\bibinfo{author}{\bibfnamefont{J.}~\bibnamefont{James}},
  \bibinfo{journal}{Quant. Fin.} \textbf{\bibinfo{volume}{3}},
  \bibinfo{pages}{114} (\bibinfo{year}{2003}).

\bibitem[{\citenamefont{Zhang}(1999)}]{ZMEM}
\bibinfo{author}{\bibfnamefont{Y.-C.} \bibnamefont{Zhang}},
  \bibinfo{journal}{Physica A} \textbf{\bibinfo{volume}{269}},
  \bibinfo{pages}{30} (\bibinfo{year}{1999}).

\bibitem[{\citenamefont{Shleifer}(2003)}]{Shleifer}
\bibinfo{author}{\bibfnamefont{A.}~\bibnamefont{Shleifer}},
  \emph{\bibinfo{title}{Inefficient markets}} (\bibinfo{publisher}{Oxford
  University Press}, \bibinfo{address}{New-York}, \bibinfo{year}{2003}).

\bibitem[{\citenamefont{Campbell et~al.}(1996)\citenamefont{Campbell, Lo, and
  MacKinlay}}]{LoBook}
\bibinfo{author}{\bibfnamefont{J.~Y.} \bibnamefont{Campbell}},
  \bibinfo{author}{\bibfnamefont{A.~W.} \bibnamefont{Lo}}, \bibnamefont{and}
  \bibinfo{author}{\bibfnamefont{A.~C.} \bibnamefont{MacKinlay}},
  \emph{\bibinfo{title}{The Econometrics of Financial Markets}}
  (\bibinfo{publisher}{Princeton University Press}, \bibinfo{year}{1996}).

\bibitem[{\citenamefont{Dacorogna et~al.}(2001)\citenamefont{Dacorogna, Gençay,
  Müller, Olsen, and Pictet}}]{Daco}
\bibinfo{author}{\bibfnamefont{M.~M.} \bibnamefont{Dacorogna}},
  \bibinfo{author}{\bibfnamefont{R.}~\bibnamefont{Gençay}},
  \bibinfo{author}{\bibfnamefont{U.~A.} \bibnamefont{Müller}},
  \bibinfo{author}{\bibfnamefont{R.~B.} \bibnamefont{Olsen}}, \bibnamefont{and}
  \bibinfo{author}{\bibfnamefont{O.~V.} \bibnamefont{Pictet}},
  \emph{\bibinfo{title}{An Introduction to High-Frequency Finance}}
  (\bibinfo{publisher}{Academic Press}, \bibinfo{address}{London},
  \bibinfo{year}{2001}).

\bibitem[{\citenamefont{Bouchaud and Potters}(2000)}]{BouchaudPottersBook}
\bibinfo{author}{\bibfnamefont{J.-P.} \bibnamefont{Bouchaud}} \bibnamefont{and}
  \bibinfo{author}{\bibfnamefont{M.}~\bibnamefont{Potters}},
  \emph{\bibinfo{title}{Theory of Financial Risks}}
  (\bibinfo{publisher}{Cambridge University Press, Cambridge},
  \bibinfo{year}{2000}).

\bibitem[{\citenamefont{Arthur et~al.}(1997)\citenamefont{Arthur, Holland,
  LeBaron, Palmer, and Tayler}}]{SantaFe}
\bibinfo{author}{\bibfnamefont{W.~B.} \bibnamefont{Arthur}},
  \bibinfo{author}{\bibfnamefont{J.~H.} \bibnamefont{Holland}},
  \bibinfo{author}{\bibfnamefont{B.}~\bibnamefont{LeBaron}},
  \bibinfo{author}{\bibfnamefont{R.}~\bibnamefont{Palmer}}, \bibnamefont{and}
  \bibinfo{author}{\bibfnamefont{P.}~\bibnamefont{Tayler}}, in
  \emph{\bibinfo{booktitle}{The Economy as an Evolving Complex System, II}},
  edited by \bibinfo{editor}{\bibfnamefont{W.~B.} \bibnamefont{Arthur}},
  \bibinfo{editor}{\bibfnamefont{S.}~\bibnamefont{Durlauf}}, \bibnamefont{and}
  \bibinfo{editor}{\bibfnamefont{D.}~\bibnamefont{Lane}}
  (\bibinfo{publisher}{Addison-Wesley}, \bibinfo{year}{1997}), vol.
  \bibinfo{volume}{XXVII} of \emph{\bibinfo{series}{SFI Studies in the Sciences
  of Complexit}}.

\bibitem[{\citenamefont{Sato and Takayasu}(1998)}]{Takayasu}
\bibinfo{author}{\bibfnamefont{A.-H.} \bibnamefont{Sato}} \bibnamefont{and}
  \bibinfo{author}{\bibfnamefont{H.}~\bibnamefont{Takayasu}},
  \bibinfo{journal}{Physica A} \textbf{\bibinfo{volume}{250}},
  \bibinfo{pages}{231} (\bibinfo{year}{1998}).

\bibitem[{\citenamefont{Lux and Marchesi}(1999)}]{Lux}
\bibinfo{author}{\bibfnamefont{T.}~\bibnamefont{Lux}} \bibnamefont{and}
  \bibinfo{author}{\bibfnamefont{M.}~\bibnamefont{Marchesi}},
  \bibinfo{journal}{Nature} \textbf{\bibinfo{volume}{397}},
  \bibinfo{pages}{498} (\bibinfo{year}{1999}).

\bibitem[{\citenamefont{Cont and Bouchaud}(2000)}]{ContBouch}
\bibinfo{author}{\bibfnamefont{R.}~\bibnamefont{Cont}} \bibnamefont{and}
  \bibinfo{author}{\bibfnamefont{J.-P.} \bibnamefont{Bouchaud}},
  \bibinfo{journal}{Macroecon. Dyn.} \textbf{\bibinfo{volume}{4}},
  \bibinfo{pages}{170} (\bibinfo{year}{2000}).

\bibitem[{\citenamefont{Challet and Zhang}(1997)}]{CZ97}
\bibinfo{author}{\bibfnamefont{D.}~\bibnamefont{Challet}} \bibnamefont{and}
  \bibinfo{author}{\bibfnamefont{Y.-C.} \bibnamefont{Zhang}},
  \bibinfo{journal}{Physica A} \textbf{\bibinfo{volume}{246}},
  \bibinfo{pages}{407} (\bibinfo{year}{1997}).

\bibitem[{\citenamefont{Slanina and Zhang}(2001)}]{SZ99}
\bibinfo{author}{\bibfnamefont{F.}~\bibnamefont{Slanina}} \bibnamefont{and}
  \bibinfo{author}{\bibfnamefont{Y.-C.} \bibnamefont{Zhang}},
  \bibinfo{journal}{Physica A} \textbf{\bibinfo{volume}{289}},
  \bibinfo{pages}{290} (\bibinfo{year}{2001}).

\bibitem[{\citenamefont{Johnson et~al.}(2000)\citenamefont{Johnson, Hart, Hui,
  and Zheng}}]{J99}
\bibinfo{author}{\bibfnamefont{N.~F.} \bibnamefont{Johnson}},
  \bibinfo{author}{\bibfnamefont{M.}~\bibnamefont{Hart}},
  \bibinfo{author}{\bibfnamefont{P.~M.} \bibnamefont{Hui}}, \bibnamefont{and}
  \bibinfo{author}{\bibfnamefont{D.}~\bibnamefont{Zheng}},
  \bibinfo{journal}{ITJFA} \textbf{\bibinfo{volume}{3}} (\bibinfo{year}{2000}).

\bibitem[{\citenamefont{Jefferies et~al.}(2001)\citenamefont{Jefferies, Hart,
  Hui, and Johnson}}]{J00}
\bibinfo{author}{\bibfnamefont{P.}~\bibnamefont{Jefferies}},
  \bibinfo{author}{\bibfnamefont{M.}~\bibnamefont{Hart}},
  \bibinfo{author}{\bibfnamefont{P.}~\bibnamefont{Hui}}, \bibnamefont{and}
  \bibinfo{author}{\bibfnamefont{N.}~\bibnamefont{Johnson}},
  \bibinfo{journal}{Eur. Phys. J. B} \textbf{\bibinfo{volume}{20}},
  \bibinfo{pages}{493} (\bibinfo{year}{2001}).

\bibitem[{\citenamefont{Challet et~al.}(2001)\citenamefont{Challet, Marsili,
  and Zhang}}]{CMZ01}
\bibinfo{author}{\bibfnamefont{D.}~\bibnamefont{Challet}},
  \bibinfo{author}{\bibfnamefont{M.}~\bibnamefont{Marsili}}, \bibnamefont{and}
  \bibinfo{author}{\bibfnamefont{Y.-C.} \bibnamefont{Zhang}},
  \bibinfo{journal}{Physica A} \textbf{\bibinfo{volume}{294}},
  \bibinfo{pages}{514} (\bibinfo{year}{2001}).

\bibitem[{\citenamefont{Challet and Marsili}(2003)}]{CM03}
\bibinfo{author}{\bibfnamefont{D.}~\bibnamefont{Challet}} \bibnamefont{and}
  \bibinfo{author}{\bibfnamefont{M.}~\bibnamefont{Marsili}},
  \bibinfo{journal}{Phys. Rev. E} \textbf{\bibinfo{volume}{68}},
  \bibinfo{pages}{036132} (\bibinfo{year}{2003}).

\bibitem[{\citenamefont{Bouchaud et~al.}(2001)\citenamefont{Bouchaud, Giardina,
  and M\'ezard}}]{BouchMG}
\bibinfo{author}{\bibfnamefont{J.-P.} \bibnamefont{Bouchaud}},
  \bibinfo{author}{\bibfnamefont{I.}~\bibnamefont{Giardina}}, \bibnamefont{and}
  \bibinfo{author}{\bibfnamefont{M.}~\bibnamefont{M\'ezard}},
  \bibinfo{journal}{Quant. Fin.} \textbf{\bibinfo{volume}{1}},
  \bibinfo{pages}{212} (\bibinfo{year}{2001}).

\bibitem[{\citenamefont{Giardina and Bouchaud}(2002)}]{BouchaudGiardina}
\bibinfo{author}{\bibfnamefont{I.}~\bibnamefont{Giardina}} \bibnamefont{and}
  \bibinfo{author}{\bibfnamefont{J.-P.} \bibnamefont{Bouchaud}}
  (\bibinfo{year}{2002}).

\bibitem[{\citenamefont{Andersen and Sornette}(2003)}]{Sornette}
\bibinfo{author}{\bibfnamefont{J.~V.} \bibnamefont{Andersen}} \bibnamefont{and}
  \bibinfo{author}{\bibfnamefont{D.}~\bibnamefont{Sornette}},
  \bibinfo{journal}{Eur. Phys. J. B} \textbf{\bibinfo{volume}{31}},
  \bibinfo{pages}{141} (\bibinfo{year}{2003}),
  \bibinfo{note}{cond-mat/0205423}.

\bibitem[{\citenamefont{Marsili}(2001)}]{M01}
\bibinfo{author}{\bibfnamefont{M.}~\bibnamefont{Marsili}},
  \bibinfo{journal}{Physica A} \textbf{\bibinfo{volume}{299}},
  \bibinfo{pages}{93} (\bibinfo{year}{2001}).

\bibitem[{\citenamefont{Martino et~al.}(2003)\citenamefont{Martino, Giardina,
  and Mosetti}}]{Andrea}
\bibinfo{author}{\bibfnamefont{A.~D.} \bibnamefont{Martino}},
  \bibinfo{author}{\bibfnamefont{I.}~\bibnamefont{Giardina}}, \bibnamefont{and}
  \bibinfo{author}{\bibfnamefont{G.}~\bibnamefont{Mosetti}},
  \bibinfo{journal}{J. Phys. A: Math. Gen.} \textbf{\bibinfo{volume}{36}},
  \bibinfo{pages}{8935} (\bibinfo{year}{2003}).

\bibitem[{\citenamefont{Kozlowski and Marsili}(2003)}]{MarsiliMaj}
\bibinfo{author}{\bibfnamefont{P.}~\bibnamefont{Kozlowski}} \bibnamefont{and}
  \bibinfo{author}{\bibfnamefont{M.}~\bibnamefont{Marsili}},
  \bibinfo{journal}{JPHA} \textbf{\bibinfo{volume}{36}}, \bibinfo{pages}{11725}
  (\bibinfo{year}{2003}).

\bibitem[{\citenamefont{Challet}()}]{ChalletPatterns}
\bibinfo{author}{\bibfnamefont{D.}~\bibnamefont{Challet}}.

\bibitem[{\citenamefont{Luce}(1959)}]{Logit}
\bibinfo{author}{\bibfnamefont{R.~D.} \bibnamefont{Luce}},
  \emph{\bibinfo{title}{Individual Choice Behavior: A Theoretical Analysis}}
  (\bibinfo{publisher}{Wiley}, \bibinfo{year}{1959}).

\bibitem[{\citenamefont{Marsili and Challet}(2001)}]{MC00}
\bibinfo{author}{\bibfnamefont{M.}~\bibnamefont{Marsili}} \bibnamefont{and}
  \bibinfo{author}{\bibfnamefont{D.}~\bibnamefont{Challet}},
  \bibinfo{journal}{Adv. Complex Systems} \textbf{\bibinfo{volume}{3}},
  \bibinfo{pages}{3} (\bibinfo{year}{2001}).

\bibitem[{\citenamefont{Farmer}(2002)}]{Farmer}
\bibinfo{author}{\bibfnamefont{J.~D.} \bibnamefont{Farmer}},
  \bibinfo{journal}{Industial and Corporate Change}
  \textbf{\bibinfo{volume}{11}}, \bibinfo{pages}{895} (\bibinfo{year}{2002}).

\bibitem[{\citenamefont{Cavagna}(1999)}]{Cavagna}
\bibinfo{author}{\bibfnamefont{A.}~\bibnamefont{Cavagna}},
  \bibinfo{journal}{Phys. Rev. E} \textbf{\bibinfo{volume}{59}},
  \bibinfo{pages}{R3783} (\bibinfo{year}{1999}).

\bibitem[{\citenamefont{Lamper et~al.}(2002)\citenamefont{Lamper, Howison, and
  Johnson}}]{Jcrash}
\bibinfo{author}{\bibfnamefont{D.}~\bibnamefont{Lamper}},
  \bibinfo{author}{\bibfnamefont{S.}~\bibnamefont{Howison}}, \bibnamefont{and}
  \bibinfo{author}{\bibfnamefont{N.}~\bibnamefont{Johnson}},
  \bibinfo{journal}{Phys. Rev. Lett.} \textbf{\bibinfo{volume}{88}},
  \bibinfo{pages}{017902} (\bibinfo{year}{2002}).

\bibitem[{\citenamefont{Garrahan et~al.}(2000)\citenamefont{Garrahan, Moro, and
  Sherrington}}]{Moro1}
\bibinfo{author}{\bibfnamefont{J.~P.} \bibnamefont{Garrahan}},
  \bibinfo{author}{\bibfnamefont{E.}~\bibnamefont{Moro}}, \bibnamefont{and}
  \bibinfo{author}{\bibfnamefont{D.}~\bibnamefont{Sherrington}},
  \bibinfo{journal}{Phys. Rev. E} \textbf{\bibinfo{volume}{62}},
  \bibinfo{pages}{R9} (\bibinfo{year}{2000}), \bibinfo{note}{cond-mat/0004277}.

\bibitem[{\citenamefont{Heimel and Coolen}(2001)}]{CoolenBatch}
\bibinfo{author}{\bibfnamefont{J.~A.~F.} \bibnamefont{Heimel}}
  \bibnamefont{and} \bibinfo{author}{\bibfnamefont{A.~C.~C.}
  \bibnamefont{Coolen}}, \bibinfo{journal}{Phys. Rev. E}
  \textbf{\bibinfo{volume}{63}}, \bibinfo{pages}{056121}
  (\bibinfo{year}{2001}).

\bibitem[{\citenamefont{Coolen and Heimel}(2001)}]{CoolenOnline}
\bibinfo{author}{\bibfnamefont{A.~C.~C.} \bibnamefont{Coolen}}
  \bibnamefont{and} \bibinfo{author}{\bibfnamefont{J.~A.~F.}
  \bibnamefont{Heimel}}, \bibinfo{journal}{J. Phys. A: Math. Gen.}
  \textbf{\bibinfo{volume}{34}}, \bibinfo{pages}{10783} (\bibinfo{year}{2001}).

\bibitem[{\citenamefont{Challet et~al.}(2000)\citenamefont{Challet, Marsili,
  and Zecchina}}]{CMZ99}
\bibinfo{author}{\bibfnamefont{D.}~\bibnamefont{Challet}},
  \bibinfo{author}{\bibfnamefont{M.}~\bibnamefont{Marsili}}, \bibnamefont{and}
  \bibinfo{author}{\bibfnamefont{R.}~\bibnamefont{Zecchina}},
  \bibinfo{journal}{Phys. Rev. Lett.} \textbf{\bibinfo{volume}{84}},
  \bibinfo{pages}{1824} (\bibinfo{year}{2000}).

\bibitem[{\citenamefont{Marsili and Ferreira}(2003)}]{Mcorr}
\bibinfo{author}{\bibfnamefont{M.}~\bibnamefont{Marsili}} \bibnamefont{and}
  \bibinfo{author}{\bibfnamefont{F.~F.} \bibnamefont{Ferreira}}
  (\bibinfo{year}{2003}).

\bibitem[{\citenamefont{Dominicis}(1978)}]{Dominicis}
\bibinfo{author}{\bibfnamefont{C.~D.} \bibnamefont{Dominicis}},
  \bibinfo{journal}{Phys. Rev B} \textbf{\bibinfo{volume}{18}},
  \bibinfo{pages}{4913} (\bibinfo{year}{1978}).

\bibitem[{\citenamefont{Wyart and Bouchaud}()}]{Wyart}
\bibinfo{author}{\bibfnamefont{M.}~\bibnamefont{Wyart}} \bibnamefont{and}
  \bibinfo{author}{\bibfnamefont{J.-P.} \bibnamefont{Bouchaud}},
  \bibinfo{note}{preprint http://arxiv.org/abs/cond-mat/0303584}.

\bibitem[{\citenamefont{Marsili}(2002)}]{MarsiliClust}
\bibinfo{author}{\bibfnamefont{M.}~\bibnamefont{Marsili}},
  \bibinfo{journal}{Quant. Fin.} \textbf{\bibinfo{volume}{2}}
  (\bibinfo{year}{2002}).

\bibitem[{\citenamefont{Cavagna et~al.}(1999)}]{Oxf1}
\bibinfo{author}{\bibfnamefont{A.}~\bibnamefont{Cavagna}} \bibnamefont{et~al.},
  \bibinfo{journal}{Phys. Rev. Lett.} \textbf{\bibinfo{volume}{83}},
  \bibinfo{pages}{4429} (\bibinfo{year}{1999}).

\end{thebibliography}

%\begin{references}
%\bibitem{bs}

%\vfill\eject
%\onecolumn

%\begin{figure}
%\centerline{\includegraphics[width=0.4\textwidth]{absfig1.eps}
%\caption{}
%\label{fig1}
%\end{figure}

\end{document}